\pdfoutput=1
\documentclass[twocolumn]{aastex61}

\usepackage{amsmath}
\usepackage{epstopdf}
\usepackage{longtable}
\usepackage{diagbox}

\begin{document}

\title{Model-Independent Reconstruction of the Cosmological Scale Factor as a Function of Lookback Time}

\correspondingauthor{Tong-Jie Zhang}
\email{tjzhang@bnu.edu.cn}

\author{Jian-Chen Zhang}
\affil{Department of Astronomy, Beijing Normal University Beijing, 100875, China \\}
\affiliation{College of Computer and Information Engineering, Dezhou University, Dezhou 253023, China \\}

\author{Jing Zheng}
\affil{Department of Astronomy, Beijing Normal University Beijing, 100875, China \\}
\affiliation{Department of Physics, Washington University in St. Louis, St. Louis, MO 63130, USA \\}


\author{Tong-Jie Zhang}
\affiliation{Department of Astronomy, Beijing Normal University Beijing, 100875, China \\}
\affiliation{Institute for Astronomical Science, Dezhou University, Dezhou 253023, China \\}



\begin{abstract}
We present a model-independent method of reconstructing scale factor against lookback time from the Observational Hubble parameter Data (OHD). The reconstruction method is independent of dynamical models and is only based on the Friedmann-Robertson-Walker metric. We also calculate the propagation of error in the reconstruction process. The reconstruction data errors mainly come from trapezoidal rule approximation and the uncertainty from OHD. Furthermore, the model discrimination ability of original OHD and reconstructed $a\text{-} t$ data is discussed under a dimensionless standard method. $a \text{-} t$ data can present the differences between cosmology models more clearly than $H\text{-}z$ data by comparing their coefficients of variations. Finally, we add fifty simulated $H(z)$ data to estimate the influence of future observation. More Hubble measurements in the future will help constrain cosmological parameters more accurately.
\end{abstract}





\section{Introduction} \label{sec:intro}
The cosmological scale factor $a(t)$ is one of the most fundamental quantities that describe the smooth background universe, although it is not observable. A \textbf{whole} relation between scale factor $a$ and cosmic time $t$ almost contains all information about cosmological kinematics, such as the expansion history, the Hubble parameter (expansion rate) $H =  \dot{a} / a$ ($\dot{a} \equiv da / dt$) and the deceleration parameter $q = -a \ddot{a} / \dot{a}^2$. \citet{2014AJ....148...94R} introduced a model-independent method of developing a scale factor - lookback time plot from Type Ia supernovae (SNe Ia) and radio-galaxy data, i.e., the Hubble diagram of modulus data against redshift. They used the scale factor plot as a better way to find the transition redshift of the universe at which the universe transitions from decelerating to accelerating.

The assumption that cosmic curvature equals zero was made in the reconstruction process from Type Ia supernovae to scale factor - lookback time data \citep{2004ApJ...612..652D}. We reconstruct the scale factor against cosmic time from the Observational Hubble parameter Data (OHD). The expression of the lookback time contains an integral of Hubble parameter, so using OHD to reconstruct cosmic time is model-independent and does not require any assumptions. We use the trapezoidal rule for approximating integrals of Hubble parameters. The reconstruction data errors come from the trapezoidal rule and OHD's errors.

The scale factor $a$ - cosmic time $t$ has a more basic status than Hubble parameter $H$ - redshift $z$, since the former directly appears in the Friedmann-Robertson-Walker metric. Although the error propagation from OHD to reconstructed data would increase errors, we expect that the $a \text{-} t$ data have a higher sensitivity to model discrimination. We calculate the $H \text{-} z$ and its variance as well as $a \text{-} t$ for several models such as Phantom, $\Lambda$CDM, Chevallier-Polarski-Linder (CPL) parametrization. The result shows $a \text{-} t$ plot has better model discrimination than $H \text{-} z$ plot by comparing the coefficient of variation, which although can not be seen directly from the figure.

The primary purpose of this paper is to present a model-independent approach to reconstruct the scale factor against cosmic lookback time data from OHD. The paper is organized as follows. We introduce the reconstruction method in section \ref{sec:reconMethod}. The error propagation is also calculated there. The reconstructed $a \text{-} t$ data are show\textbf{n} in section \ref{sec:avstData}. Section \ref{sec:ModelSel} discusses the model discrimination ability of original $H \text{-} z$ and reconstructed $a \text{-} t$ data. In the section \ref{sec:sim}, we use simulated $H(z)$ data to forecast the improvement effects of future $H(z)$ observation on this reconstruction.

\section{Reconstruction}
\label{sec:reconMethod}

\subsection{Dimensionless Cosmic Time $\tau$}
\label{sec:tDef}
The Friedmann-Robert-Walker metric is
\begin{equation}
	ds^2 = c^2 dt^2 - a^2(t) \left( \frac{dr^2}{1 - kr^2} + r^2 d\theta^2 + r^2 \sin^2 \theta d\varphi^2 \right)
\label{equ:frwMetric}
\end{equation}
where $t$ is cosmological time. The age of the universe corresponding to redshift $z$ is
\begin{equation}
	t(z) = t_H \int_z^{+ \infty} \frac{dz'}{(1 + z') E(z')}.
\label{equ:tz}
\end{equation}
where $t_H \equiv 1/H_0$, $H_0$ is the Hubble constant and $t_H$ is called Hubble time. $E(z) = H(z) / H_0$.

The upper limit of the integral of $t(z)$ is infinity, whereas observation data only exist in low redshift ($0 \le z \le 2.4$ for the data we use). To solve this problem, we turn to the lookback time $t_L$, which is the time measured back from the present epoch $t_0$ to any earlier time $t(z)$. It can be written as
\begin{equation}
\label{equ:lookbacktime}
	t_L (z) = t_0 - t(z) = t_H \int_0^z \frac{dz'}{(1 + z')\, E(z')}.
\end{equation}

The ``direction'' of $t_L$ is opposite from $t(z)$. For simplicity, we define \textit{dimensionless cosmic time} $\tau$ by Hubble time $t_H$ as:
\begin{equation}
\label{equ:tDef}
	\tau \equiv 1 - \frac{t_L}{t_H} = 1 - \int_0^z \frac{dz'}{(1 + z') E(z')}.
\end{equation}
We use $\tau$ as \textbf{a} reconstructed quantity that replaces the ``true'' cosmic time $t(z)$.

\begin{figure}[ht!]
	\plotone{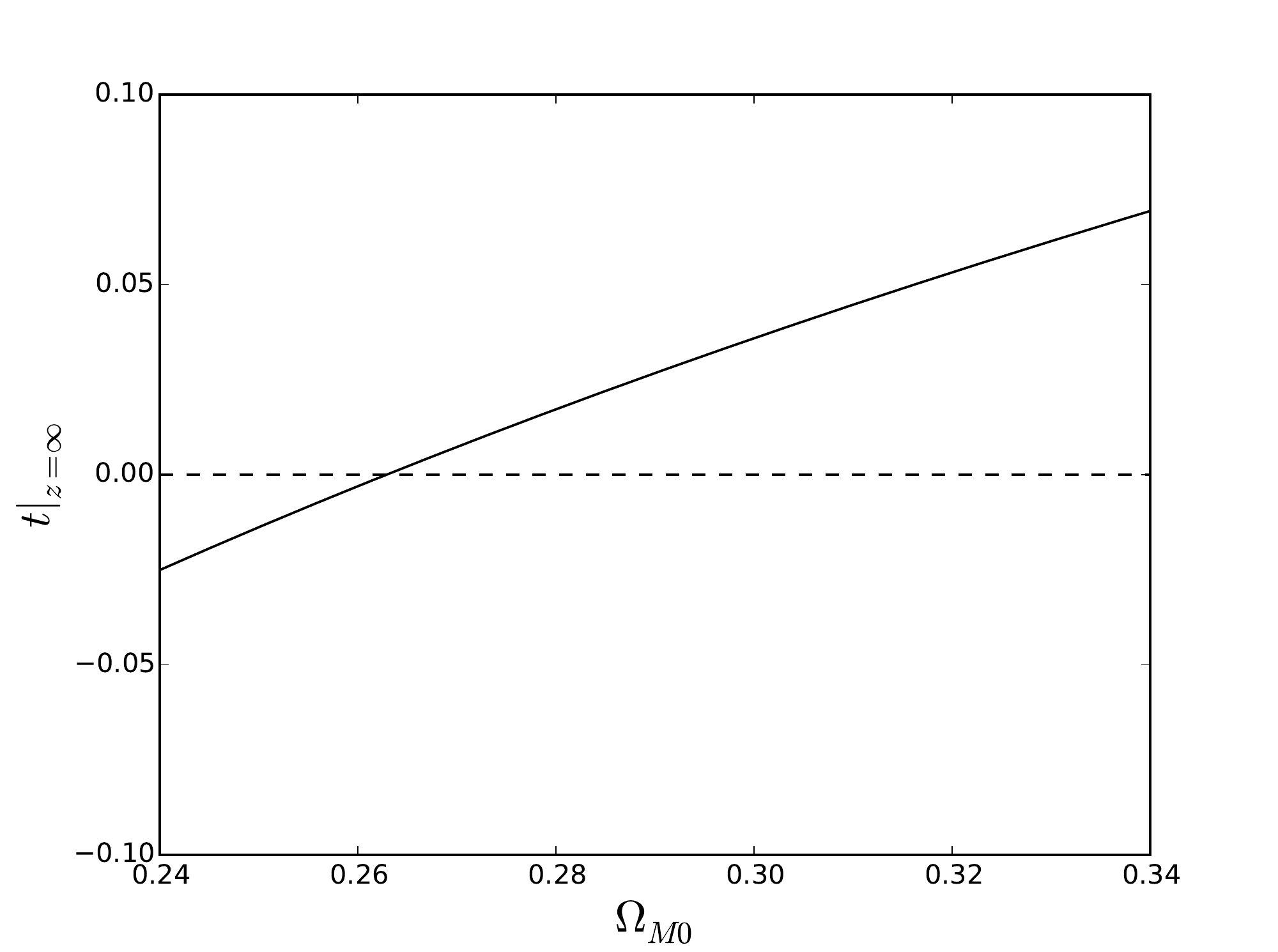}
	\caption{$\tau|_{z \rightarrow \infty}$ (see equation \eqref{equ:tauzinf}) in $\Lambda$CDM model with varing parameter $\Omega_m$. The range of variation is $[0.24, 0.34]$, and the latest fitting result of $\Omega_m$ from Planck is $0.316 \pm 0.014$ \citep{2020A&A...641A...6P}. From the figure we can see that $\tau|_{z \rightarrow \infty} \approx 0$ in our universe.
		\label{fig:tauvars}}
\end{figure}

It is obvious that $\tau|_{z = 0} = 1$, i.e., $\tau = 1$ in \text{the} present epoch, but $\tau$ is subtly different from $t$. For example, when $z \to +\infty$(the Big Bang),
\begin{equation}
	\tau(z \to +\infty) = 1 - \int_0^{+\infty} \frac{dz'}{(1 + z') E(z')} \neq 0, \quad \text{in general.} \label{equ:tauzinf}
\end{equation}
$\tau$ in Big Bang is not zero. In $\Lambda$CDM model, $\tau(z \to +\infty)$ depends on the cosmological density parameters by the expansion rate $E(z)$. But according to current observation constraint values, $\tau|_{z \rightarrow \infty} \approx 0$ \citep{2020A&A...641A...6P}, see Figure \ref{fig:tauvars}.

The relation between scale factor $a$ and dimensionless cosmic time $\tau$ can be expressed as
\begin{equation}
    \tau = 1 + \int_1^a \frac{1}{a' \cdot E(a')} \, da' \ ,
\end{equation}
where we use $a = a_0 / (1 + z)$ and set $a_0 = 1$.

\subsection{Reconstruction Method}
\label{sec:atRecs}
The OHD contains
\begin{equation}
	\label{equ:OHDdef}
		\{ z_i, H_i, \Delta H_i \}, \quad i = 1, \dots, N.
\end{equation}
where $\Delta H_i$ means the error of OHD. ``$\Delta$'' is used to represent data error.

From the chain rule and the relation $a=1/(1+z)$, we can get:
\begin{equation}
	\label{equ:dt_da}
		\frac{d \tau}{dz} = \frac{d \tau}{da} \frac{d a}{d z} = -a^2 \frac{d \tau}{d a}.
\end{equation}
The process that starts from OHD to get $\left( a_i, (d\tau / da)_i \right)$ is:
\begin{eqnarray}
		\mathrm{OHD}: (z_i,\, H_i) \stackrel{\mathrm{dimensionless}} {\Longrightarrow}& &(z_i,\, E_i) \\
		\stackrel{\mathrm{equation} \eqref{equ:tDef}}{\Longrightarrow} & &\left[ z_i,\, \left( \frac{d \tau}{dz} \right)_i \right] \\
		\stackrel{\mathrm{equation} (\eqref{equ:dt_da})\   \mathrm{and} \ a=1/(1+z)}{\Longrightarrow} & &\left[ a_i,\, -a_i^2 \left( \frac{d \tau}{da} \right)_i \right] \\
		\Longrightarrow &&\left[ a_i,\, \left( \frac{d \tau}{da} \right)_i \right].
\end{eqnarray}
Next, we reconstruct $(a_i, \tau_i)$ from $\displaystyle{ \left( a_i,\, \left( \frac{d \tau}{d a} \right)_i \right) }$. According to $\tau_0 = 1, a_0 = 1$ (present value), we integrate the derivative:
\begin{eqnarray}
		\int_a^{a_0 = 1} \left( \frac{d \tau}{da} \right) da = \int_\tau^{\tau_0 = 1 } d\tau \\
		\tau = 1 - \int_a^1  \left( \frac{d \tau}{da} \right) da \ .
\end{eqnarray}
For simplicity, we define auxiliary quantity $s$ by:
\begin{equation}
	s \equiv \int_a^1 \left( \frac{d \tau}{da} \right) da \ ,
\end{equation}
so $\tau = 1 - s$. We only have discrete data points $\{ a_i, \big( d\tau/da \big)_i \}$, so we use trapezoidal rule to approximate integrations:	
\begin{eqnarray}
	s_1 &\approx& \left( \frac{d \tau}{d a} \right)_1 (1 - a_1) \ , \notag \\
	s_2 &\approx& s_1 + \frac{1}{2} \left[ \left( \frac{d \tau}{d a} \right)_1 + \left( \frac{d \tau}{d a} \right)_2 \right] (a_1 - a_2) \ , \notag \\
	s_3 &\approx& s_2 + \frac{1}{2} \left[ \left( \frac{d \tau}{d a} \right)_2 + \left( \frac{d \tau}{d a} \right)_3 \right] (a_2 - a_3) \ , \notag \\
	& \cdots & \notag \\
	s_n &\approx& s_{n-1} + \frac{1}{2} \left[ \left( \frac{d \tau}{d a} \right)_{n-1} + \left( \frac{d \tau}{d a} \right)_{n} \right] (a_{n-1} - a_{n}) \ , \notag \\
	\tau_i &=& 1 - s_i,  \quad i = 1, \dots , n.
\end{eqnarray}
In this way, the $\{a_i, \tau_i \}$ data are finally reconstructed.

\subsection{Error Propagation}
\label{sec:errpro}
It's easily to get $\{a_i,\ (d \tau/da)_i,\ \Delta(d \tau/da)_i \}$ from $\{z_i,\ H_i,\ \Delta H_i\}$, because both of them are related to cosmic expansion `velocity'. The corresponding error propagation is:
\begin{eqnarray}
	\Delta E_i &=& E_i \cdot \left( \frac{\Delta H_i}{ H_i } + \frac{\Delta H_0}{H_0} \right) \\
	\Delta \left(\frac{d \tau}{dz} \right)_i &=& \left(\frac{d \tau}{dz} \right)_i \cdot \frac{\Delta E_i}{E_i} \\
	\Delta \left( \frac{d \tau}{da} \right)_i &=& \left( \frac{d \tau}{da} \right)_i \cdot \frac{\Delta \left(\frac{d \tau}{dz} \right)_i}{\left(\frac{d \tau}{dz} \right)_i} \ ,
\end{eqnarray}
where the Hubble constant $H_0 = 67.4\  \pm\  0.5 \,\mathrm{km\ s^{-1} Mpc^{-1}}$, which comes from Planck 2018 data \citep{2020A&A...641A...6P}.

The process from $d \tau/da$ to $\tau$ causes most errors since we use trapezoidal rule approximates an integral. Trapezoidal method has its own proper error, and $d\tau / da$ data errors also contribute,
\begin{equation}
	\begin{split}
	s_n  &=  \int_{a_n}^1 \left(\frac{d \tau}{da} \right) da \\ &\approx  \left(\frac{d \tau}{da}\right)_1 (1 - a_1) \\
	& \quad + \frac{1}{2} \sum_{i=2}^n \left[\left(\frac{d \tau}{da}\right)_{i-1} + \left(\frac{d \tau}{da}\right)_{i} \right] (a_{i-1} - a_i).
	\end{split}
\end{equation}
and
\begin{equation}
	\Delta s_n  =  \Delta s_{n, \text{trape}} + \Delta s_{n, \text{data}}.
\end{equation}

Errors of $s$ that correspond to errors of $d \tau/da$ are easily calculated:
\begin{equation}
\label{equ:dsdata}
\begin{split}
	\Delta s_{n, \text{data}} &= \Delta \left(\frac{d \tau}{da}\right)_1 (1 - a_1) \\ &+ \frac{1}{2} \sum_{i=2}^n \left[\Delta \left(\frac{d \tau}{da}\right)_{i-1} + \Delta  \left(\frac{d \tau}{da}\right)_{i} \right] (a_{i-1} - a_i).
\end{split}
\end{equation}

We estimate trapezoidal rule's proper error by:
\begin{equation}
\label{equ:trapeErr}
	\left| \int_a^b f(x) dx - \frac{b-a}{2} [f(a) + f(b)] \right| \le \frac{1}{4} M (b - a)^2,
\end{equation}
where constant $M$ satisfies $|f'(x)| \le M,\, \forall x$.

The maximum of $(d\tau / da)$ reconstruction data is $1.5$, and we take $M = 10$ for an estimate of trapezoidal rule error:
\begin{eqnarray}
	\Delta s_{1, \text{trape}} & = & \frac{1}{4} M (1 - a_1)^2 \\
\label{equ:dstrape}
	\Delta s_{n, \text{trape}} & = & \Delta s_1 + \sum_{i=2}^n \frac{1}{4} M (a_{i-1} - a_{i})^2 \\
	\Delta \tau_i & = & \Delta s_i, \quad i = 1, \dots, n.
\end{eqnarray}

\section{Reconstruction Results}
\label{sec:avstData}

\begin{longtable}{cccc}
\caption{The 43 $H(z)$ measurements contains redshift z, Hubble parameters $H(z)$(km $\rm s^{-1}$ $\rm Mpc^{-1}$) and the uncertainties of Hubble parameters $\sigma_{H}$(km $\rm s^{-1}$ $\rm Mpc^{-1}$).\label{tab:OHDdata}}\\
\hline
\emph{z} & $H(z)$ & $\sigma_{H}$  & References \\
\hline
\endfirsthead
0.09	&	$	69$&$12	$	 &	\citet{2003ApJ...593..622J} \\
\hline
0.17	&	$	83$&$8	$&	\\
0.27	&	$	77$&$14	$	 &	\\
0.4	&	$	95$&$17	$	 &	 \\
0.9	&	$	117$&$23	$	 &	\citet{2005PhRvD..71l3001S} \\
1.3	&	$	168$&$17	$	&	\\
1.43	&	$	177$&$18	$	 &	\\
1.53	&	$	140$&$14	$ &	\\
1.75	&	$	202$&$40	$ &	\\
\hline
0.48	&	$	97$&$62	$ &	\citet{2010JCAP...02..008S} \\
0.88	&	$	90$&$40	$	 &	\\
\hline
0.1791	&	$	75$&$4	$ &	\\
0.1993	&	$	75$&$5	$ &	\\
0.3519	&	$	83$&$14	$	 &	\\
0.5929	&	$	104$&$13	$ &	\citet{2012JCAP...07..053M} \\
0.6797	&	$	92$&$8	$ &	\\
0.7812	&	$	105$&$12	$	 &	\\
\hline
\emph{z} & $H(z)$ & $\sigma_{H}$  & References \\
\hline
0.8754	&	$	125$&$17	$ &	 \citet{2012JCAP...07..053M} \\
1.037	&	$	154$&$20	$ &	\\
\hline
0.07	&	$	69$&$19.6	$	 &	\\
0.12	&	$	68.6$&$26.2	$ &	\citet{2014RAA....14.1221Z} \\
0.2	&	$	72.9$&$29.6	$	 &	\\
0.28	&	$	88.8$&$36.6	$	 &	\\
\hline
1.363	&	$	160$&$33.6	$	 &	\citet{2015MNRAS.450L..16M} \\
1.965	&	$	186.5$&$50.4	$&	\\
\hline
0.3802	&	$	83$&$13.5	$ &	\\
0.4004	&	$	77$&$10.2	$ &	\\
0.4247	&	$	87.1$&$11.2	$ &	\citet{2016JCAP...05..014M} \\
0.4497	&	$	92.8$&$12.9	$ &	\\
0.4783	&	$	80.9$&$9	$ &	\\
\hline
0.47   &      $      89$&$34        $      &   \citet{2017MNRAS.467.3239R} \\
\hline
0.24	&	$	79.69$&$2.65	$	&	\citet{2009MNRAS.399.1663G} \\
0.43	&	$	86.45$&$3.68	$	 &	\\
\hline
0.44	&	$	82.6$&$7.8	$	&	\\
0.6	&	$	87.9$&$6.1	$	&	\citet{2012MNRAS.425..405B}\\
0.73	&	$	97.3$&$7	$	&	\\
\hline
0.35   &      $      82.1$&$4.9     $      &   \citet{2012MNRAS.426..226C} \\
\hline
0.35	&	$	84.4$&$7	$ &	\citet{2013MNRAS.431.2834X}\\
\hline
0.57	&	$	92.4$&$4.5	$ &	\citet{2013MNRAS.429.1514S}\\
\hline
2.3	&	$	224$&$8	$ &\cite{2013AA...552A..96B}\\
\hline
0.57    &      $      92.9$&$7.8$     &\cite{2014MNRAS.439...83A}\\
\hline
2.36&$226$&$8	$&	\citet{2014JCAP...05..027F}\\
\hline
2.34	&	$	222$&$7	$&\citet{2015AA...574A..59D} \\
\hline
\end{longtable}

\begin{figure}[htp!]
	\plotone{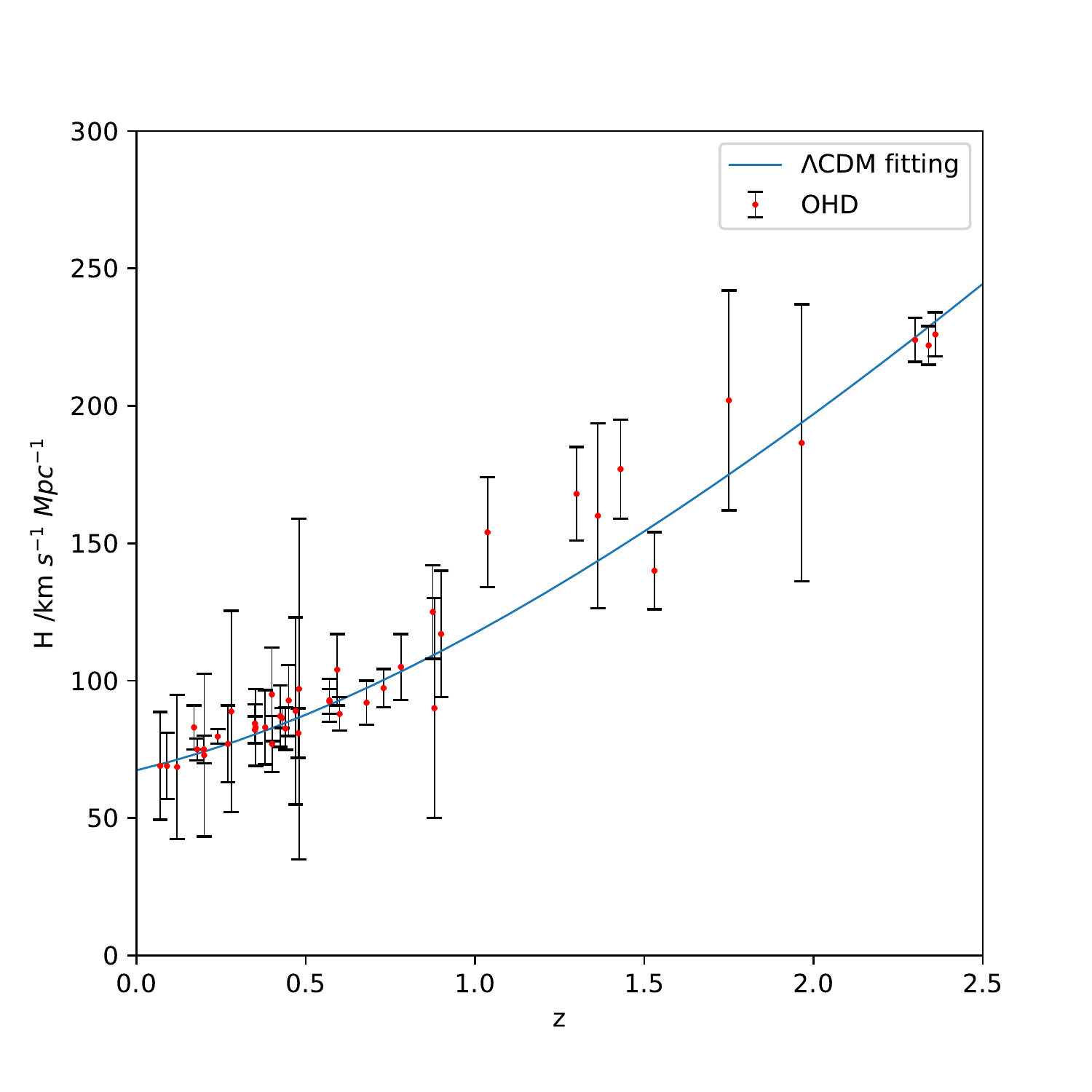}
	\caption{43 Hubble measurements(red points with error bars) and the  $\Lambda$CDM model (blue line) fitted from 43 OHD. The fitting result is $\Omega_m = 0.290 \pm 0.008$. \label{fig:OHD}}
\end{figure}

\begin{figure}[ht!]
	\plotone{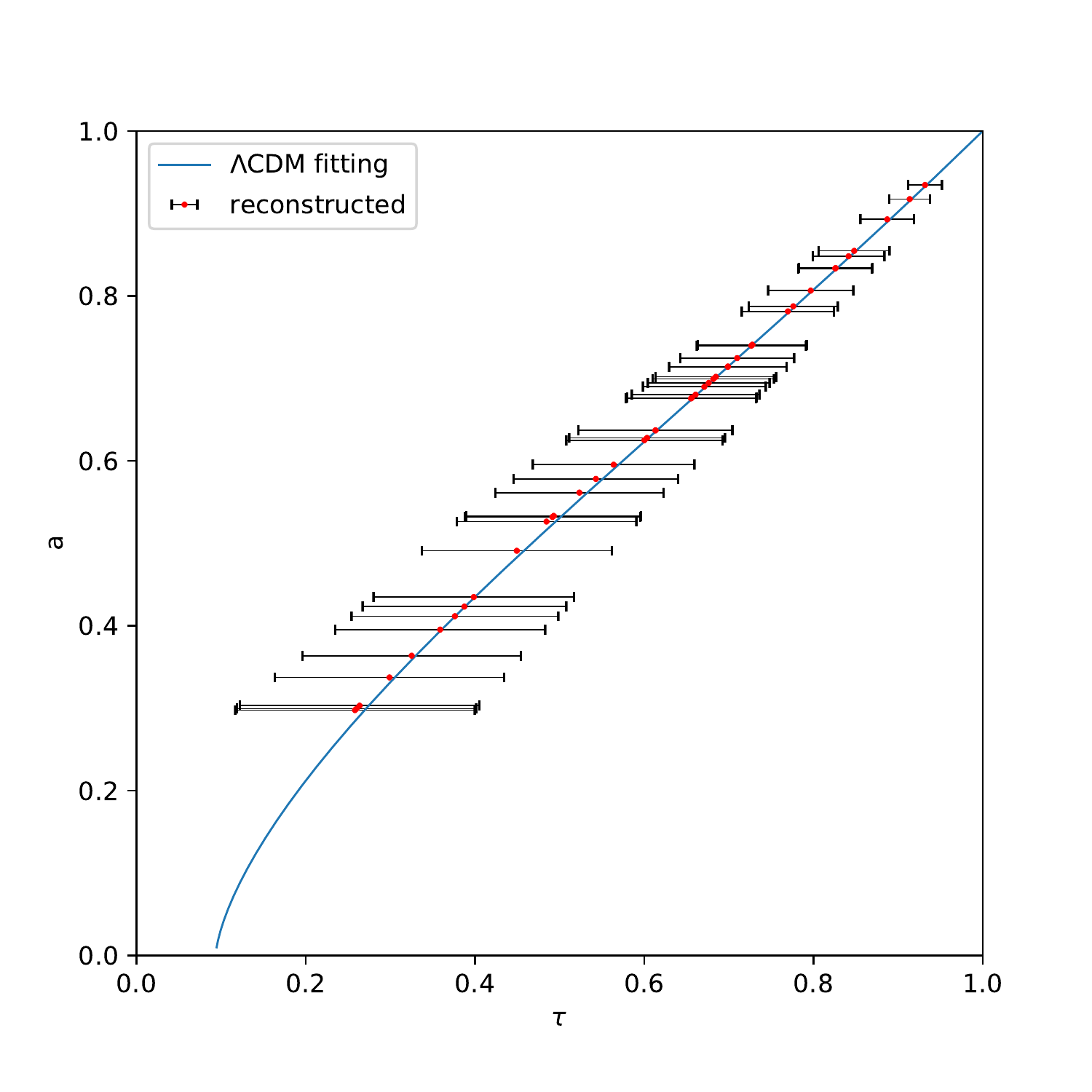}
	\caption{The reconstructed $a\text{-}\tau$ data with $\Lambda$CDM model fitting results ($\Omega_{M0} = 0.372 \pm 0.087$). Note that $a|_{\tau = 0} \neq 0$, because of the definition of $\tau$(equation \eqref{equ:tDef}), $\tau|_{z \rightarrow \infty} = 1 - t_0/t_H$.}
	\label{fig:dataFit}
\end{figure}

The 43 OHD are shown in Figure \ref{fig:OHD} and Table \ref{tab:OHDdata}, which contains 31 Hubble measurements from the method of cosmic chronometers and 12 from the radial BAO method. The reconstructed $a \text{-} d \tau/da$ and $a\text{-} \tau$ data are shown in Figure \ref{fig:dataFit} and Table \ref{tab:data}. The error in the early universe (low $\tau$) is enormous. According to equation \eqref{equ:dsdata} and \eqref{equ:dstrape}, the error is accumulated in calculating high-$z$ or low-$\tau$ data. The main sources of error are the accumulation effect of trapezoidal rule and errors of Hubble measurements. In order to reduce the error from trapezoidal rule, more future OHD data is needed to increase the number of approximate trapezoids to have a more accurate result.

To test the validation of reconstruction data, we use Markov Chain Monte Carlo(MCMC) method to fit the matter density parameter $\Omega_m$ in $\Lambda$CDM model:
\begin{eqnarray}
	E(z) &=& \sqrt{\Omega_m (1 + z)^3 + \Omega_\Lambda} \\
	E(a) &=& \sqrt{\frac{\Omega_m}{a^3} + \Omega_\Lambda} \\
	\Omega_m &+& \Omega_\Lambda = 1
\end{eqnarray}
where we ignore radiation ($\Omega_{R}$) and consider a flat model ($\Omega_k = 0$).

According to equation \eqref{equ:tDef} and $a = 1 / (1 + z)$, the relation between $\tau$ and $a$ in $\Lambda$CDM is:
\begin{eqnarray}
	\tau(a) &=& 1 + \int_1^{a} \frac{da'}{a' \cdot E(a')}
\end{eqnarray}
where the only free parameter is $\Omega_m$.


\begin{deluxetable}{c|cc|ccc}
	\tablecaption{The reconstructed $a\text{-}d\tau/da$ and $a\textrm{-}\tau$ data. \label{tab:data}}
	\tablecolumns{6}
	\tablehead{
		\colhead{$a = 1/(1+z)$} &
		\colhead{$\displaystyle{\frac{d\tau}{d a}}$} &
		\colhead{$ \Delta \left(\displaystyle{\frac{d\tau}{d a}}\right)$} &
		\colhead{$\tau$} &
		\colhead{$\Delta \tau$} &
		\colhead{$\dfrac{\Delta \tau}{\tau} / \% $}
}
\startdata
0.9346	&1.051 	&0.311 	&0.931 	&0.020 & 2.18\\
0.9174 	&1.071 	&0.198 	&0.913 	&0.025 & 2.70\\
0.8929 	&1.107 	&0.435  &0.886 	&0.032 & 3.66\\
0.8547 	&0.956 	&0.103 	&0.847 	&0.043 & 5.05\\
0.8482 	&1.066 	&0.069 	&0.840 	&0.043 & 5.15\\
0.8338 	&1.084 	&0.085 	&0.825 	&0.044 & 5.38\\
0.8333 	&1.116 	&0.466 	&0.824 	&0.045 & 5.40\\
0.8065 	&1.055 	&0.047 	&0.795 	&0.051 & 6.47\\
0.7874 	&1.118 	&0.216 	&0.774 	&0.054 & 6.97\\
0.7813 	&0.977 	&0.414 	&0.768 	&0.056 & 7.28\\
0.7407 	&1.115 	&0.079 	&0.726 	&0.066 & 9.08\\
0.7407 	&1.084 	&0.102 	&0.726 	&0.066 & 9.08\\
0.7397 	&1.104 	&0.199 	&0.724 	&0.066 & 9.11\\
0.7245 	&1.127 	&0.196 	&0.708 	&0.069 & 9.75\\
0.7143	&0.999 	&0.190 	&0.697 	&0.071 & 10.19\\
0.7140	&1.233 	&0.177 	&0.697 	&0.071 & 10.20\\
0.7019	&1.109 	&0.155 	&0.682 	&0.073 & 10.71\\
0.6993 	&1.122 	&0.060 	&0.679 	&0.073 & 10.89\\
0.6944 	&1.182 	&0.125 	&0.674 	&0.074 & 10.95\\
0.6898	&1.059 	&0.159 	&0.668 	&0.074 & 11.14\\
0.6803	&1.120 	&0.441 	&0.658 	&0.077 & 11.75\\
0.6765	&1.239 	&0.152 	&0.654 	&0.078 & 12.00\\
0.6757 	&1.034 	&0.673 	&0.653 	&0.079 & 12.07\\
0.6369 	&1.152 	&0.069 	&0.610 	&0.093 & 15.26\\
0.6369	&1.146 	&0.109 	&0.610 	&0.093 & 15.26\\
0.6277 	&1.039 	&0.142 	&0.600 	&0.094 & 15.70\\
0.6250 	&1.234 	&0.100 	&0.597 	&0.095 & 15.84\\
0.5953 	&1.238 	&0.122 	&0.561 	&0.098 & 17.46\\
0.5780 	&1.205 	&0.100 	&0.539 	&0.100 & 18.51\\
0.5614 	&1.150 	&0.145 	&0.520 	&0.102 & 19.60\\
0.5333 	&1.017 	&0.150 	&0.489 	&0.106 & 21.67\\
0.5319 	&1.416 	&0.646 	&0.488 	&0.106 & 21.85\\
0.5263 	&1.101 	&0.229 	&0.481 	&0.109 & 22.67\\
0.4909 	&0.897 	&0.127 	&0.445 	&0.115 & 25.89\\
0.4348 	&0.928 	&0.104 	&0.394 	&0.122 & 30.90\\
0.4232 	&1.001 	&0.222 	&0.383 	&0.124 & 32.30\\
0.4115 	&0.931 	&0.105 	&0.372 	&0.126 & 33.79\\
0.3953 	&1.225 	&0.136 	&0.354 	&0.128 & 36.01\\
0.3636 	&0.923 	&0.193 	&0.320 	&0.133 & 41.46\\
0.3373 	&1.078 	&0.304 	&0.294 	&0.140 & 47.42\\
0.3030 	&0.999 	&0.047 	&0.258 	&0.145 & 56.27\\
0.2994 	&1.020 	&0.044 	&0.255 	&0.145 & 57.14\\
0.2976 	&1.008 	&0.047 	&0.253 	&0.145 & 57.59\\
\enddata
\end{deluxetable}

The result of direct fitting from OHD data is
\begin{equation}
\Omega_m = 0.290 \pm 0.008,
\end{equation}
which is close to the result from \citet{2020A&A...641A...6P} $\Omega_m = 0.315 \pm 0.007$ with similar uncertainty, but \textbf{they are} not consistent.

The fitting result from $a\text{-}\tau$ data is:
\begin{equation}
\Omega_m = 0.372 \pm 0.087.
\end{equation}
This result implies an universe with $\Lambda$-dominated accelerating expansion, which is consistent with SNe Ia and BAO results. Compared with OHD fit of $\Lambda$CDM result $\Omega_m = 0.290 \pm 0.008$, this exercise shows the reconstructed $a \text{-} \tau$'s parameter is consistent with the Planck result, which verifies the feasibility of the reconstruct\textbf{ion} method. However, it also shows its limitation that the trapezoidal rule introduces a much larger uncertainty.

\section{Model Discrimination}
\label{sec:ModelSel}
To test the $H \text{-} z$ and $a \text{-} \tau$ plot's ability of model discrimination, we find the models which are in good agreement with the results obtained from the observations, such as the Planck result \citep{2020A&A...641A...6P}, Supernovae Ia (SNe Ia), Baryon Acoustic Oscillations (BAO) and Cosmic Microwave Background (CMB) \citep{2016ApJ...821...60W}. A non-parametric smoothing method --- the Gaussian process is also used to get an independent result. The models we choose to distinguish are Phantom, $\Lambda$CDM, Chevallier-Polarski-Linder (CPL) parametrization.

The observational constraints on the parameters are $\theta_i \pm \Delta \theta_i, \ i = 1, \dots, n$, the estimated error in a model quantity $f(\theta_1, \dots, \theta_n)$ is (approximately):
\begin{equation}
	\Delta f(\theta_1, \dots, \theta_n) \approx \sqrt{ \sum_{i=1}^n \left( \frac{\partial f}{\partial \theta_i} \Delta \theta_i \right)^2}.
\end{equation}

\subsection{Phantom model}
In the Phantom model, the Hubble parameter $H$ about redshift $z$ is given by:
\begin{align}
	H_{\text{Phantom}}(z) &= H_0 \Big[ \Omega_m (1 + z)^3 + (1 - \Omega_m) (1 + z)^{3(1 + \omega)} \Big]^{1/2}.
\end{align}

The dimensionless time $\tau$ about scale factor $a$ is:
\begin{align}
	\tau_{\text{Phantom}} (a) = 1 + \int_1^a \dfrac{da'}{a' \left[ \dfrac{\Omega_m}{a'^3} +  \dfrac{1 - \Omega_m}{a'^{3(1 + \omega)}}  \right]^{1/2} }.
\end{align}
We follow \citet{2017JCAP...03..005R}'s scheme: fix $\omega = -2$ and take $\Omega_m = 0.315 \pm 0.007$ from the Planck result \citep{2020A&A...641A...6P}. Note that Phantom model with $\omega = -2$ is actually far from Planck constraints. It may not \textbf{be} a good choice to describe our real universe, but here we add this model in the hope that more different models can be compared to illustrate and test the model discrimination ability of $a \text{-} \tau$ model.

\subsection{$\Lambda$CDM model}
$\Lambda$CDM model is most widely accepted and is in very good agreement with observation. Ignoring the curve ($\Omega_k$) and radiation term ($\Omega_R$), the Hubble parameter is written as:
\begin{align}
	H_{\Lambda \text{CDM}} (z) &= H_0 \left[ \Omega_m (1+z)^3 + (1 - \Omega_m) \right]^{1/2}. \\
\intertext{The dimensionless time is:}
	\tau_{\Lambda \text{CDM}} (a) &= 1 + \int_1^a \dfrac{da'}{a' \left[ \dfrac{\Omega_m}{a'^3} + 1 - \Omega_m \right]^{1/2}}.
\end{align}
We take $\Omega_m = 0.315 \pm 0.007$ \citep{2020A&A...641A...6P}.

\subsection{Chevallier-Polarski-Linder (CPL) model}
CPL is one of the most popular parametrizations of the dark energy equation of state\citep{2001IJMPD..10..213C, 2003PhRvL..90i1301L}. It has two equation of state parameters $\omega_0, \omega_1$:
\begin{equation}
	\omega_{\text{CPL}} = \omega_0 + \omega_1 \frac{z}{1 + z} = \omega_0 + \omega_1 (1 - a).
\end{equation}
The corresponding Hubble parameter and cosmic time \textbf{are}:
\begin{align}
	H_{\text{CPL}} (z) =& H_0 \bigg[ \Omega_m (1 + z)^3 + (1 - \Omega_m) (1 + z)^{3(1+\omega_0+\omega_1)} \notag\\
	\phantom{H_{\text{CPL}} =} & \exp \left(- \frac{3\omega_1 z}{1 + z} \right) \bigg]^{1/2}. \\
	\tau_{\text{CPL}} (a) =& 1 + \notag \\
	\phantom{\tau_{\text{CPL}} (a) =}& \int_1^a \dfrac{da'}{a' \left[ \dfrac{\Omega_m}{a'^3} + \dfrac{1 - \Omega_m}{a'^{3(1 + \omega_0 + \omega_1)}} \exp \big( -3\omega_1 (1 - a') \big) \right]^{1/2}}.
\end{align}
We use $\Omega_m = 0.300 \pm 0.014, \omega_0 = -0.982 \pm 0.134, \omega_1 = -0.082^{+0.655}_{-0.440}$ obtained from the joint analysis of Supernovae Ia (SNe Ia), Baryon Acoustic Oscillation (BAO) and Cosmic Microwave Background (CMB) data \citep{2016ApJ...821...60W}.

\begin{figure}[htpb!]
\plotone{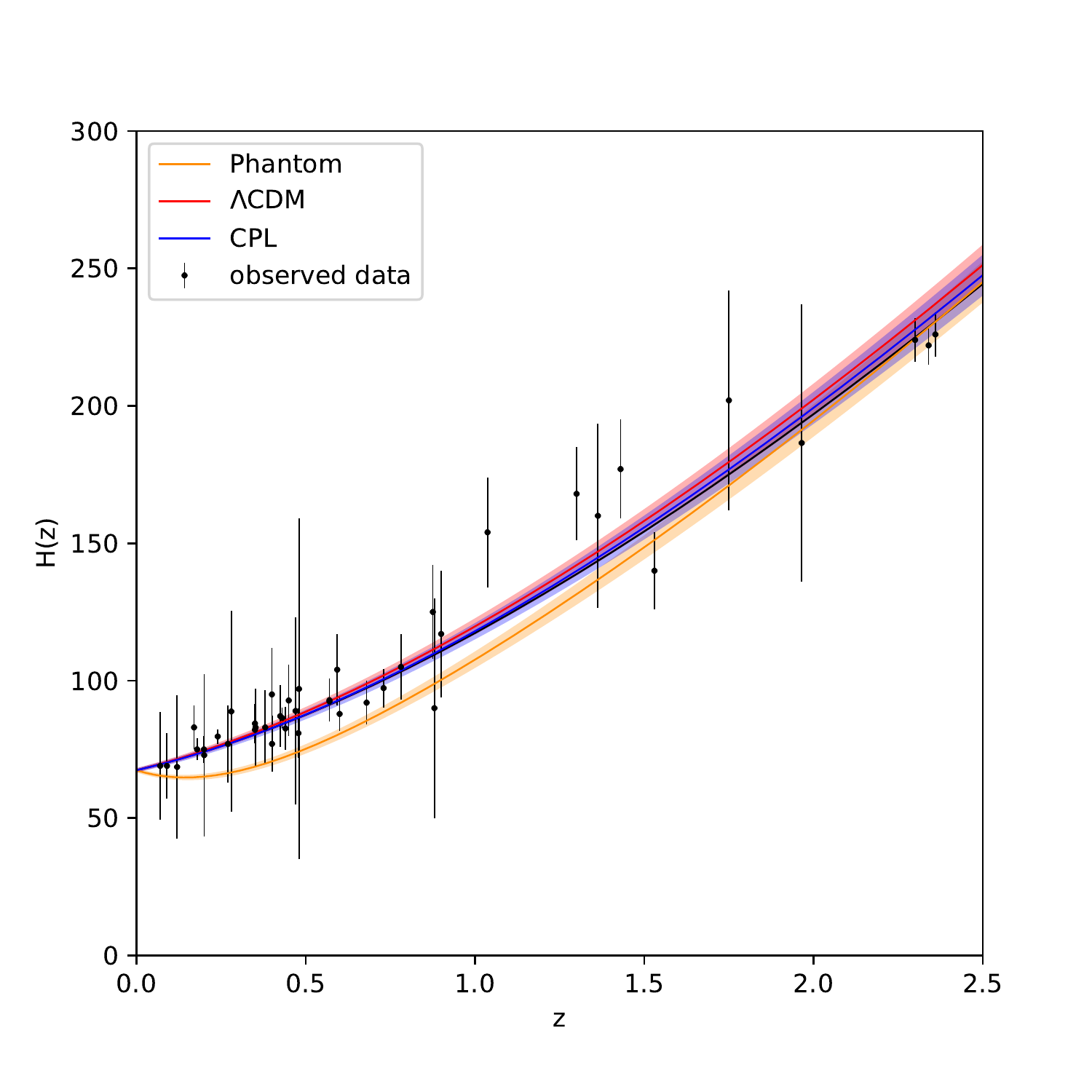}
\caption{Variation\textbf{s} of $H \text{-} z$ for Phantom, $\Lambda$CDM and CPL model\textbf{s}. \label{fig:Hzall}}
\end{figure}

\begin{figure}[htpb!]
\plotone{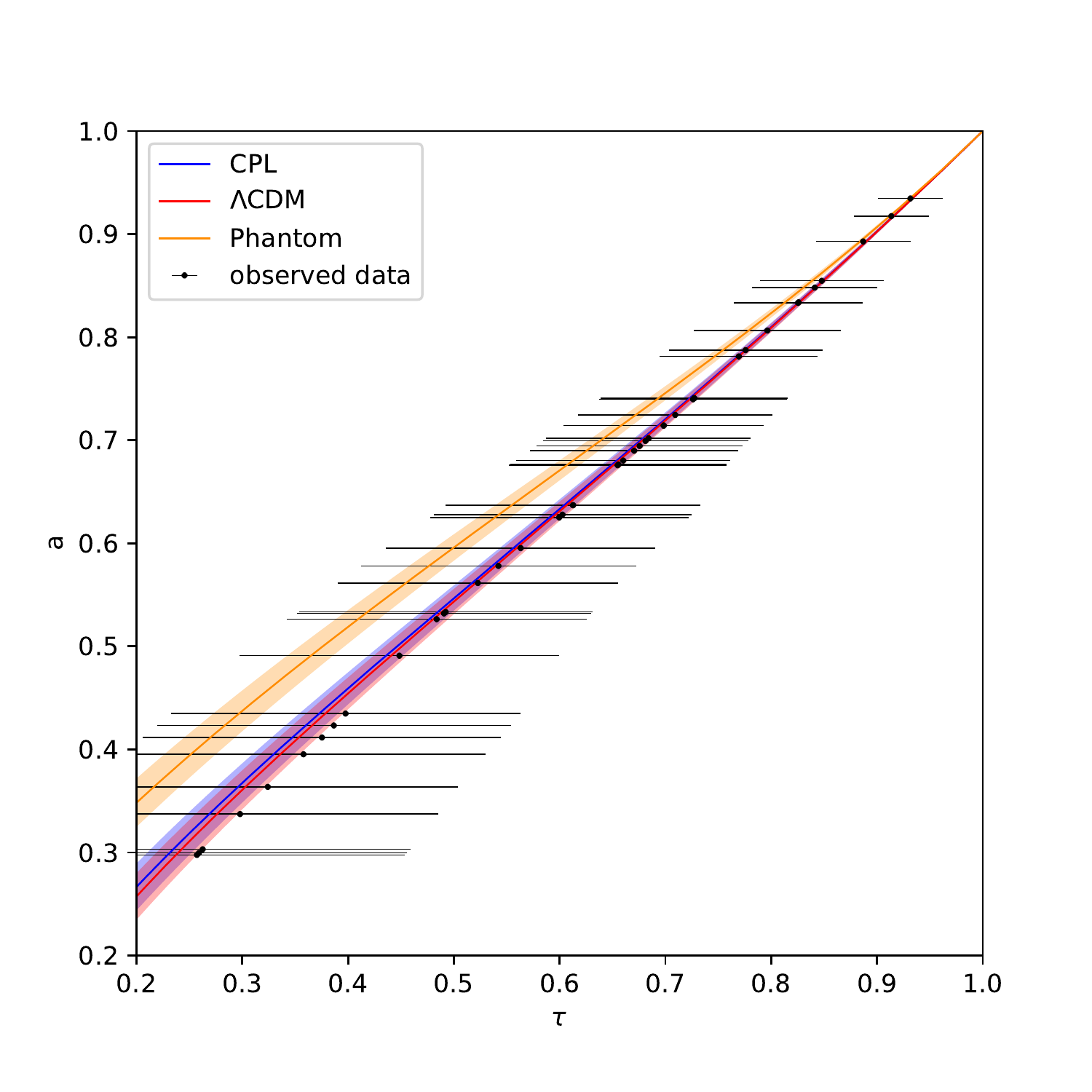}
\caption{Variation\textbf{s} of $a \text{-} \tau$ for Phantom, $\Lambda$CDM and CPL model\textbf{s}. \label{fig:atall}}
\end{figure}

The collect results of $H\text{-}z$ plot and $a \text{-} \tau$ plot  are shown in Figure \ref{fig:Hzall} and \ref{fig:atall}. In the $H\text{-}z$ plot, the curves of Phantom, $\Lambda$CDM and CPL model\textbf{s} are entangled together, and their error bands overlap, which means $H \text{-} z$ data cannot distinguish these models well. Whereas in $a \text{-} \tau$ plot, the error bands should have separated more clearly in theory due to the integral effect, but we can not see it clearly from the figure. A parameterized comparison method is needed.

Here we use the coefficient of variation(CV) to measure the dispersion of different models in a standard and dimensionless way. It is defined as the ratio of the standard deviation $\sigma$ to the average value $\mu$:
\begin{equation} \label{equ:cv}
c_v=\frac{\sigma}{\mu}
\end{equation}

The calculation result is shown in Table \ref{tab:cv}. According to the $H\text{-}z$ plot data, the differences of CV values between models are in the thousandth or less, which only account for 0.2$\%$ or less. From the CV data of the $H\text{-}z$ plot, we can see that the differences between models are in tenths or single digits, which even exceeds 100$\%$ of some CV values. Therefore, the dispersion degree between models improves by reconstructing $a \text{-} t$ data. We believe the data indicate that $a \text{-} \tau$ data's ability of model selection is better.

\begin{table}[!htbp]
\begin{center}
\caption{Coefficient of variations of different models and different plots.}
{\begin{tabular}{|c|c|c|c|}
\hline
\diagbox[dir=NW,height=20mm,width=40mm]{model}{CV}{plot} & $H\text{-}z$ & $a\text{-}\tau$ \\
\hline
Phantom & 0.699997 & 3.42210\\
\hline
$\Lambda$CDM & 0.697390 & 1.10299 \\
\hline
CPL & 0.697517 & 1.20176 \\
\hline
\end{tabular}}\label{tab:cv}
\end{center}
\end{table}

\section{Future Data Simulation}
\label{sec:sim}
In this section, we simulate future OHD to study its influence on $a\textrm{-}\tau$ method. We use the simulation method from \citet{2011ApJ...730...74M}:
\begin{equation}
H_{sim}(z)=H_{fid}(z)+\Delta H
\end{equation}
$\Delta H$ is the deviation between simulation value and fiducial value.

Data simulation has to choose a model first. We choose the $\Lambda$CDM model as our fiducial model:
\begin{eqnarray}
H_{fid}(z)=H_0\sqrt{\Omega_m(1+z)^3+\Omega_{\Lambda}}. \label{equ:fid}
\end{eqnarray}

The parameters come from \citet{2020A&A...641A...6P}, which are
\begin{eqnarray}
H_0 &=& 67.4 \pm 0.5 \ , \\
\Omega_{m} &=& 0.315 \pm 0.007 \ .
\end{eqnarray}
From the spatially flat $\Lambda$CDM model, $\Omega_{\Lambda}$ can be calculated by $\Omega_m+\Omega_{\Lambda}=1$. Then, we can obtain a set of fiducial values from equation (\ref{equ:fid}).

\begin{figure}[ht!]
	\plotone{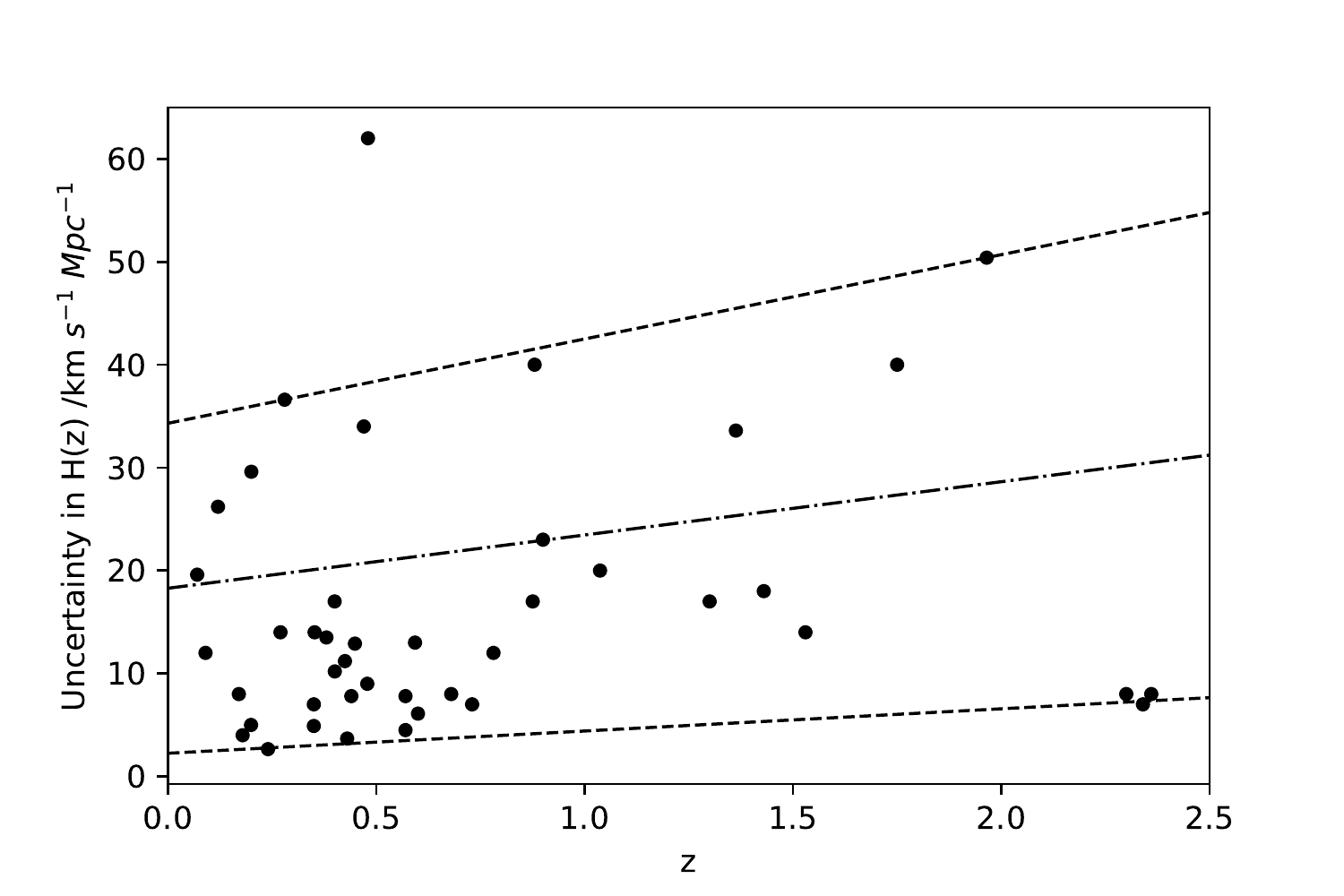}
	\caption{Uncertainties of $43$ Hubble measurements. They contain 31 Cosmic Chronometer measurements and 12 Baryon Acoustic Oscillation measurements. The upper line $\sigma_{+}$  and the lower line $\sigma_{-}$ are the boundary lines of reliable data, and the middle dotted line is the average value of two boundary lines. \label{fig:OHDsimerror}}
\end{figure}

\begin{figure}[htpb!]
	\plotone{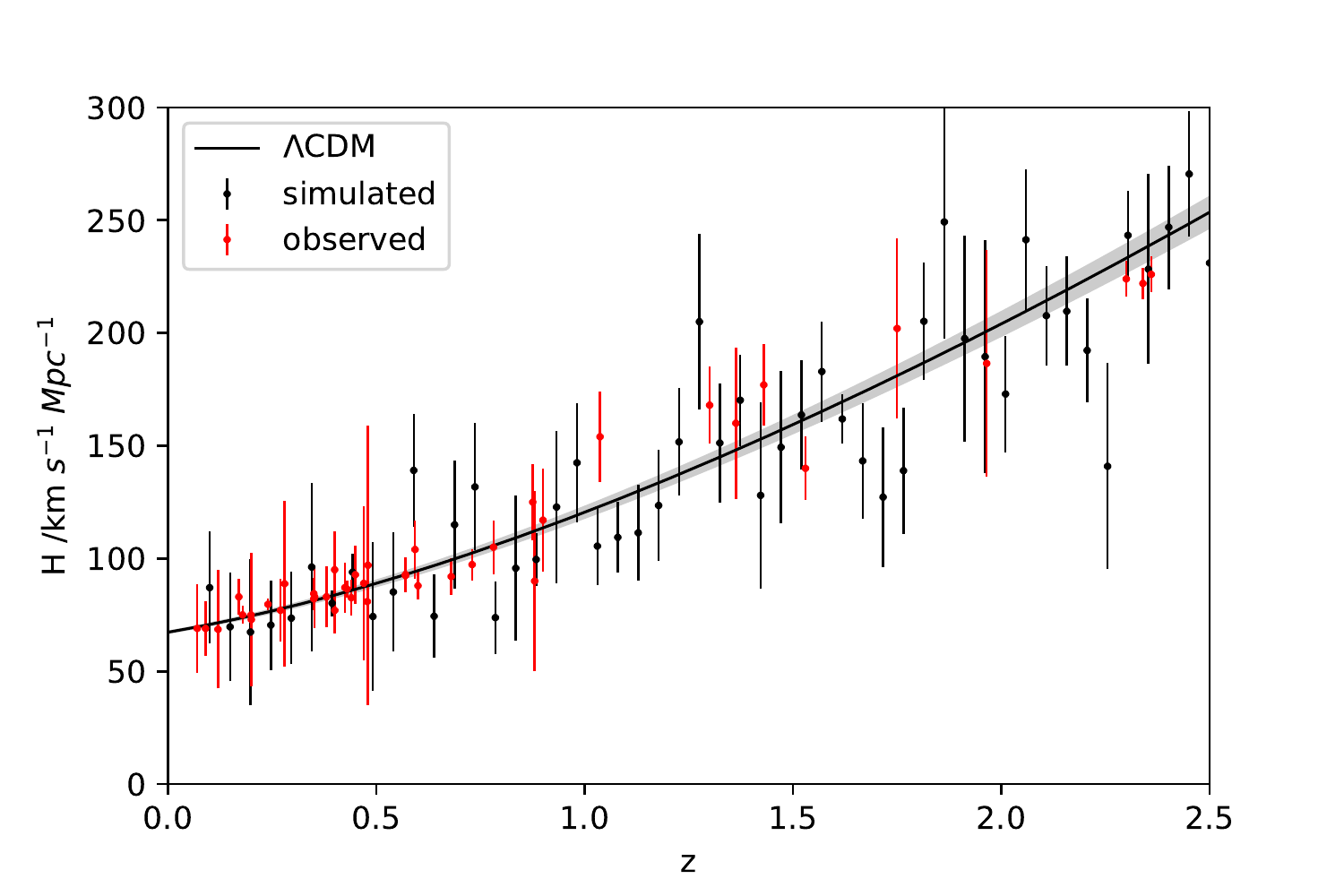}
	\caption{$50$ simulated Hubble parameters(black) with $43$ observed Hubble measurements(red). The black solid line represents the $\Lambda$CDM model and the grey region is its error band. \label{fig:OHDsim}}
\end{figure}

Next, we need to estimate the uncertainties of future OHD. 43 Hubble measurements' uncertainties are shown in Figure \ref{fig:OHDsimerror}. After removing one value with an obvious deviation, we draw two outlines to represent the general trend of the uncertainties. Two bounded lines are expressed as one upper line $\sigma_{+}=8.19z+34.31$ and one lower line $\sigma_{-}=2.16z+2.25$. This simulation method believes the future measurements will also conform to this trend, and their value will fall within this error strip between $\sigma_{-}$ and $\sigma_{+}$. In other words, an random uncertainty $\sigma(z)$ of OHD can be estimated by a Gaussian distribution $N(\sigma_0(z),(\sigma_{+}-\sigma_{-})/4)$, where $\sigma_0(z)$ is the average of $\sigma_{+}$ and $\sigma_{-}$ and the $(\sigma_{+}-\sigma_{-})$ represent $4\sigma$ trip so that the $\sigma(z)$ can fall within the strip. This random uncertainty $\sigma(z)$ can be used to determine the deviation $\Delta H$ by a Gaussian distribution $N(0,\sigma(z))$. 50 Hubble parameters are simulated through the above way, and the result is shown in Figure \ref{fig:OHDsim}.

\begin{figure}[htpb!]
	\plotone{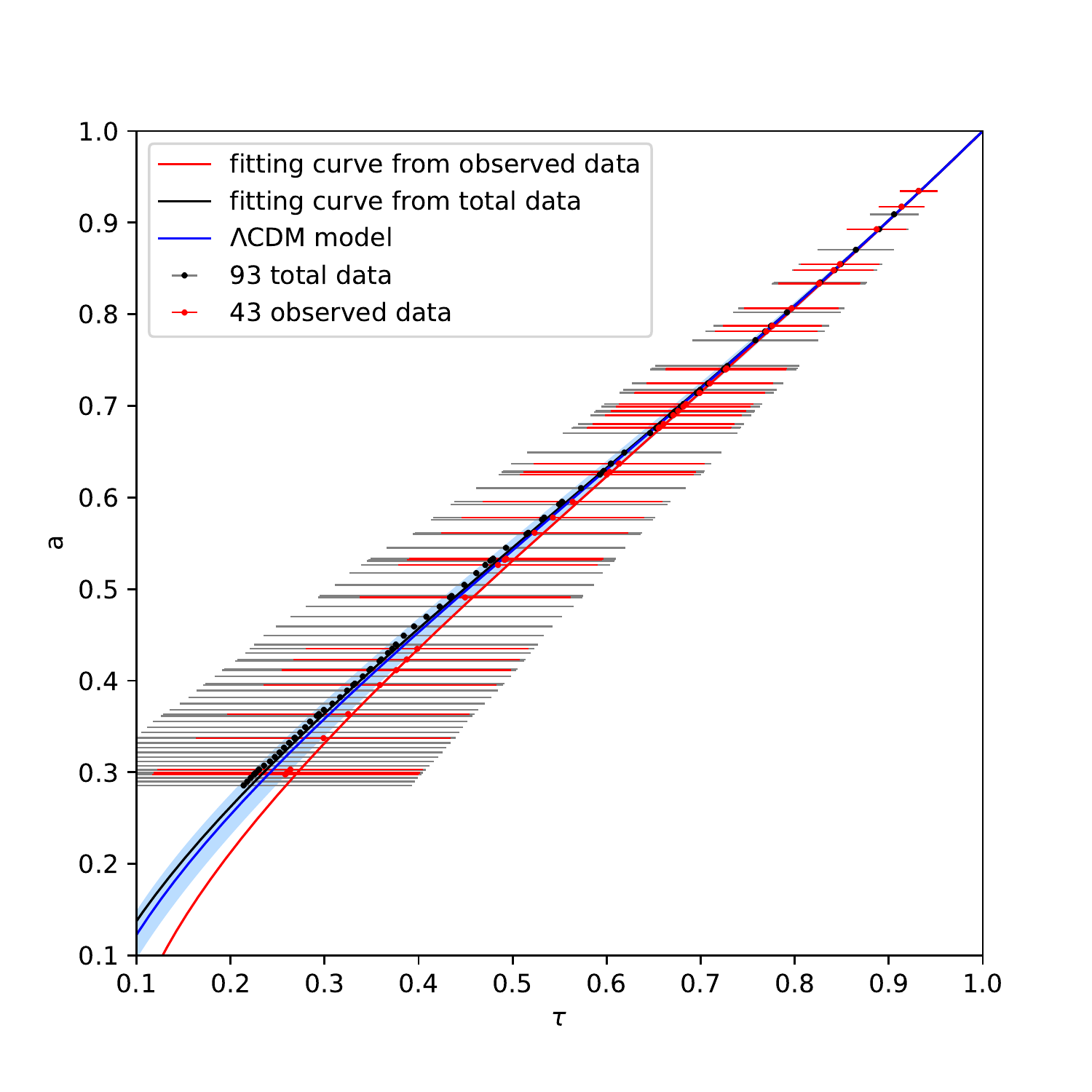}
	\caption{The reconstructed $a\text{-}\tau$ data from Hubble parameters. Red points are 43 reconstruction result from observed data. Black points are 90 reconstruction result from total OHD, which including 43 observed data and 50 simulated data. The blue curve represents the $\Lambda$CDM model from \citet{2020A&A...641A...6P}. The red curve is the best $\Lambda$CDM fitting with original 43 OHD and the black curve is the fitting with total data. Note that $a|_{\tau = 0} \neq 0$, since the definition of $\tau$(equation \eqref{equ:tDef}) implies that $\tau|_{z \rightarrow \infty} = 1 - t_0/t_H$. \label{fig:OHD_at_sim}}
\end{figure}

Through the previous reconstructed method in section \ref{sec:reconMethod} to constrain the matter density parameter $\Omega_m$, the corresponding $a\text{-}\tau$ plot is shown in Figure \ref{fig:OHD_at_sim}. When we only apply the original 43 observed data, the fitting curve describes red points well with the result of $\Omega_m = 0.372 \pm 0.087$, but it can be seen that there is still a certain gap between the fitting curve and the  $\Lambda$CDM model curve. When we apply total data, including 43 observed data and 50 simulated data, the error of data points becomes larger due to the integral effect, while the result's trend is closer to the $\Lambda$CDM model with a result of $\Omega_m = 0.303 \pm 0.047$. These are consistent with our expectations. The $a\text{-}\tau$ plot result should be more biased towards $\Lambda$CDM model after adding 50 data simulated by $\Lambda$CDM model to the sample. In addition, the error of $\Omega_m$ decreased from 0.087 to 0.047 after adding more OHD. More data points are conducive to the accuracy of the study. In another word, as more Hubble measurements are added in the future, the $a\text{-}\tau$ diagram can help to constrain cosmological parameters better and help people to study which model can describe our universe more accurately.

\section{conclusions}
We describe a model-independent method to reconstruct the scale factor $a$ against the lookback time $\tau$ from OHD. This is a new way to handle observational Hubble parameter data. 43 $H(z)$ data points are collected for reconstruction, and the redshift range is $(0.0, 2.4)$. The original OHD and reconstructed $a \text{-} \tau$ data are both plotted. The $\Lambda$CDM model presents a classic fit through the $a \text{-} \tau$ data, thus validating the reconstructed results. However, one limitation is the large error caused by the nature of the integral.

For the more fundamental position of scale factor than Hubble parameter, their abilities of model discrimination are different. By comparing the coefficient of variations of various cosmology models, we find $a \text{-} \tau$ plot can magnify the differences between models based on the integral effect so that the $a \text{-} \tau$ plot has a better model discriminate ability than the $H\text{-}z$ plot. Note that the differences are not significant enough to be observed directly from the graph. We use the coefficient of variations to compare numerically.

We simulate fifty $H(z)$ data with a fiducial $\Lambda$CDM model based on \cite{2011ApJ...730...74M} and reconstruct total $a\text{-}\tau$ data to forecast the improvement effects of future H(z) observation. On \textbf{the} one hand, due to the integration characteristics, errors are also accumulated. Therefore, when the sample takes all data, the error bar increases. On the other hand, $a\text{-}\tau$ plot shows a more accurate result of constraining cosmological parameters. If there are more Hubble measurements in the future, the $a\text{-}\tau$ method can better present the model to help people find a better model to describe our present universe.

\section*{Acknowledgements}
We sincerely appreciate the referee's suggestions, which help us greatly improve our manuscript. This work was supported by the National Science Foundation of China (NSFC) Programs Grants No. 11929301 and National Key R\&D Program of China (2017YFA0402600).
\bibliography{OHD_at_ApJ}

\end{document}